\documentclass[]{spie}  

\usepackage[]{graphicx}

\title{A method to directly image exoplanets in multi-star systems such as Alpha-Centauri} 

\author{Sandrine J. Thomas\supit{a} and Ruslan Belikov\supit{b} and Eduardo Bendek\supit{b}
\skiplinehalf
\supit{a}Large Synoptic Survey Telescope, 950N Cherry Av, Tucson AZ 85719, USA; \\
\supit{b}Nasa Ames Research Center, Moffett Field, CA 94035, USA
}

\authorinfo{Further author information: (Send correspondence to S.J.T.)\\S.J.T.: E-mail: sthomas@lsst.org, Telephone: 1 520 318 8227\\  R.B.:ruslan.belikov-1@nasa.gov, Telephone: 1 650 604 0833 }

\begin{document} 
  \maketitle 

\begin{abstract}
Direct imaging of extra-solar planets is now a reality, especially with the deployment and commissioning of the first generation of specialized ground-based instruments such as the Gemini Planet Imager and SPHERE. These systems will allow detection of Jupiter-like planets $10^7$ times fainter than their host star. Obtaining this contrast level and beyond requires the combination of a coronagraph to suppress light coming from the host star and a wavefront control system including a deformable mirror (DM) to remove residual starlight (speckles) created by the imperfections of telescope. 
However, all these current and future systems focus on detecting faint planets around single host stars, while several targets or planet candidates are located around nearby binary stars such as our neighboring star Alpha Centauri. 
Here, we present a method to simultaneously correct aberrations and diffraction of light coming from the target star as well as its companion star in order to reveal planets orbiting the target star. This method works even if the companion star is outside the control region of the DM (beyond its half-Nyquist frequency), by taking advantage of aliasing effects.  

\end{abstract}


\keywords{Wavefront Control, Binary Systems, Exoplanets, MEMS deformable mirror, space-based instrumentation,  Alpha Centauri}

\section{INTRODUCTION}
\label{sec:intro}  
The exoplanets field is rapidly expanding with the success of the Kepler mission (Burke et al. 2014 \cite{Burke14} and references therein) and the emergence of direct imaging ground based instruments (GPI \cite{Macintosh14}, SPHERE \cite{Beuzit08}, SCExAO \cite{Guyon10}, P1640 \cite{Hinkley08}). One of the most exciting prospects for future telescopes is finding other Earth analogues in our galaxy or solar neighborhood and ultimately detect life on them. The Kepler space telescope has already revealed that roughly 22\% of stars have planets between 1 and 2 Earth radii in their habitable zone \cite{Batalha14}.

However, most Sun-like stars are found in multi-star systems or visual binaries and conventional coronagraphs and wavefront control systems are designed for single-star systems, which means that they only suppress the light of the one on-axis star. Because of the perceived challenges of suppressing starlight of more than one star, multi-star systems are usually excluded from mission target lists, despite being more numerous (at least among Sun-like or earlier star systems). Therefore, there is a need for a technique enabling the direct imaging of planetary systems and disks around multi-star systems, which we present here.
In particular, our new technique potentially allows the detection of biomarkers on Earth-like planets around our nearest-neighbor star, Alpha Centauri, with a small and cheap space telescope, potentially decades sooner than a large space coronagraph could do the same around a single star system (see Belikov et al. 2015 \cite{Belikov15} and Bendek et al. 2015 \cite{Bendek15}).

 \begin{figure}[ht]
\begin{center}
\includegraphics[scale=0.3]{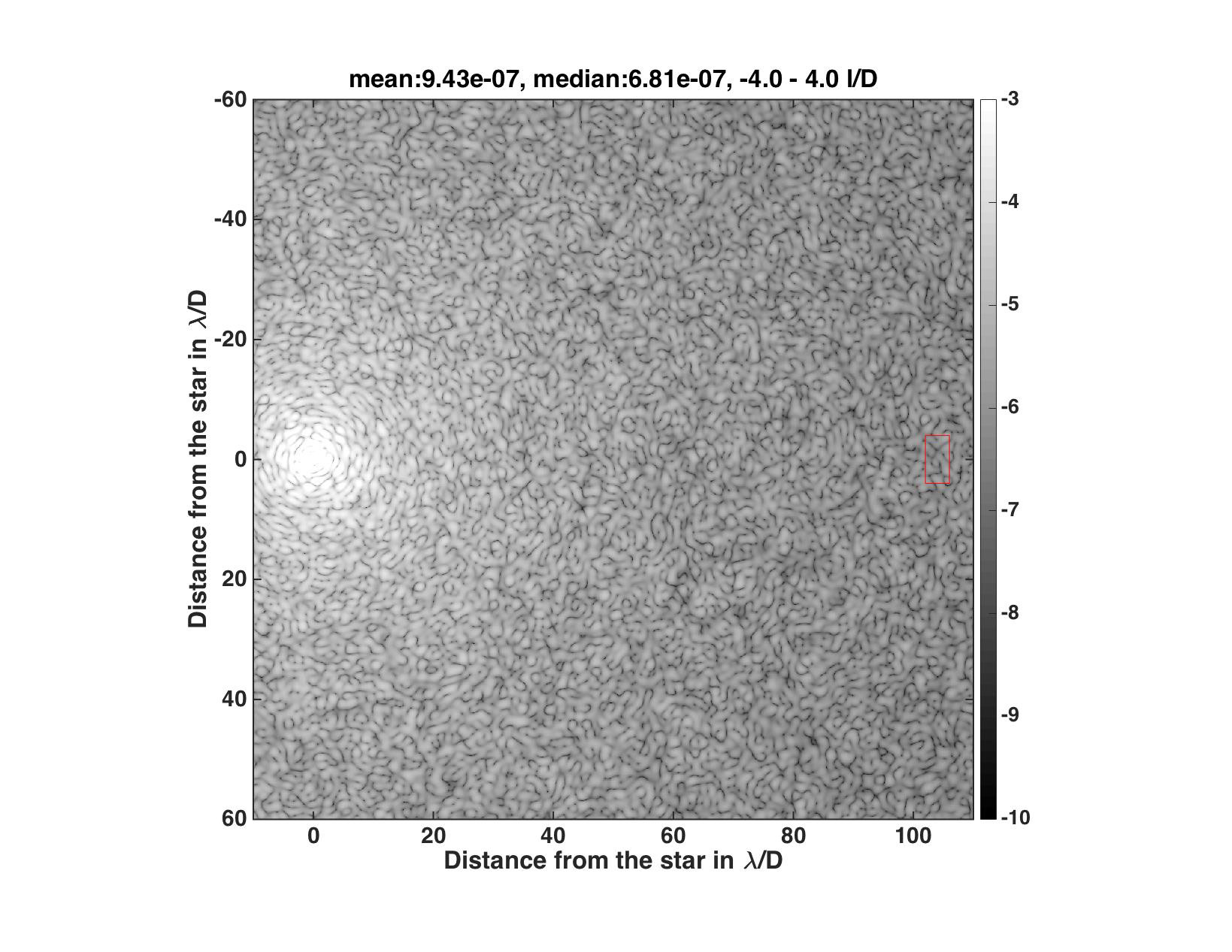}
   \includegraphics[scale=0.425]{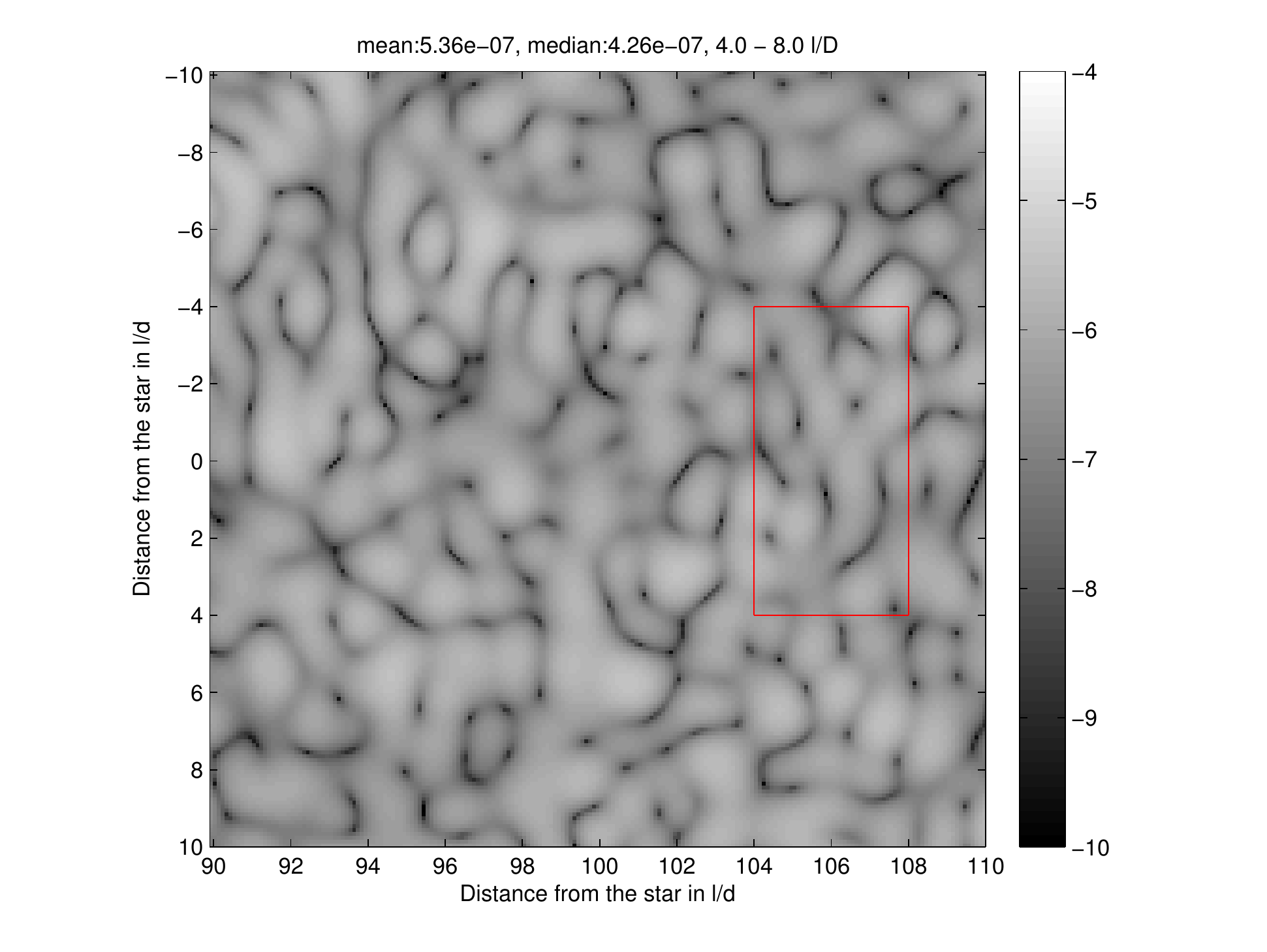}
\caption{\label{fig:issues1} Illustration of the challenges involved with creating a dark zone around a target belonging to a double star system. Left: simplified case where the on-axis star (close to the red region of interest) has been assumed to be completely suppressed, and only the light from the off-axis star (on the opposite end) is considered. Oftentimes, the region of interest around the on-axis star lies beyond the Nyquist-limited outer working angle of the deformable mirror. Right: zoomed in version of the left image, centered on the on-axis star. }
\end{center}
\end{figure}

\begin{figure}[ht]
\begin{center}
   \includegraphics[height=7cm]{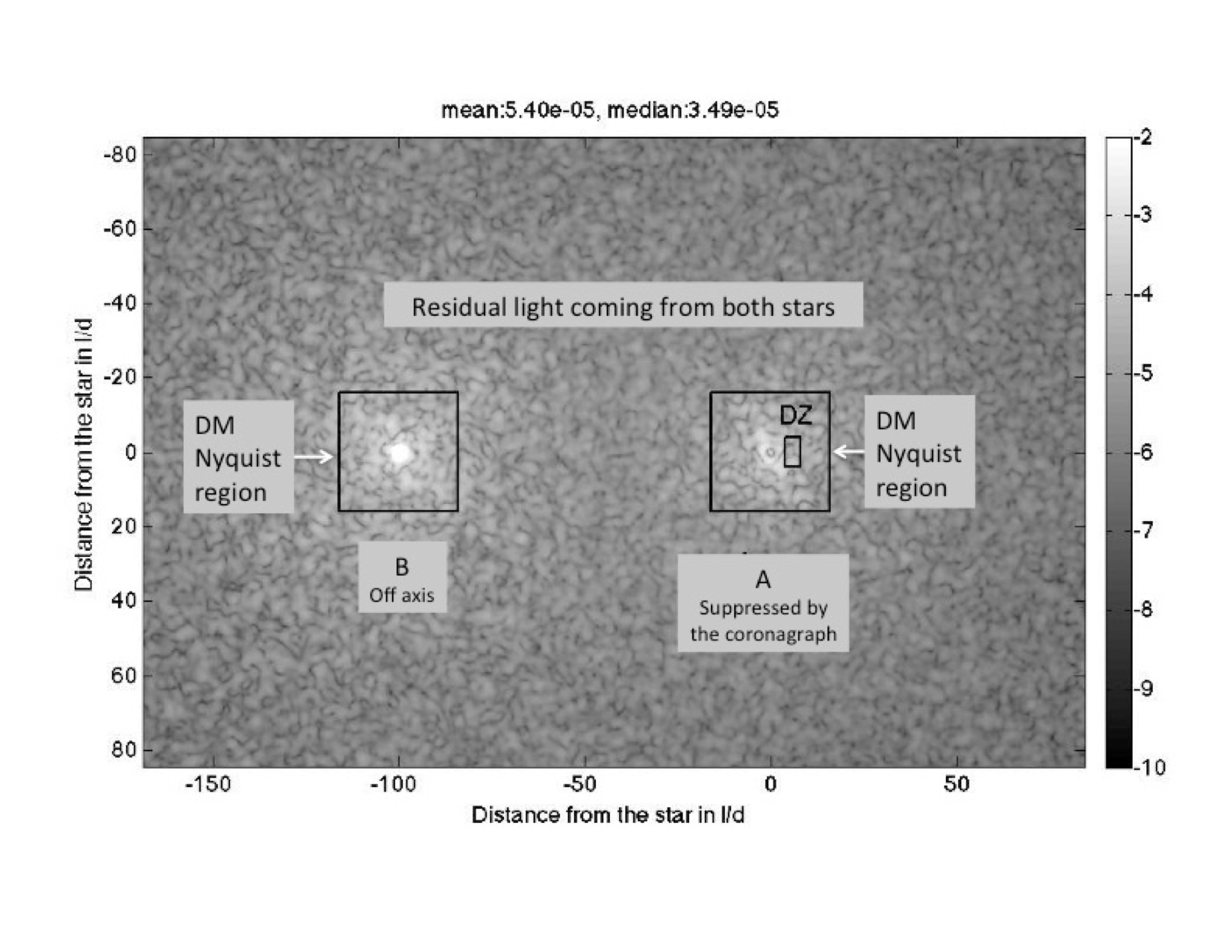}\\
\caption{\label{fig:issues2} Illustration of the challenges involved with creating a dark zone around a target belonging to a double star system. It shows a combination of light from both stars, requiring independent suppression of each, using conventional (sub-Nyquist) wavefront control for the on-axis star (A) and super-Nyquist wavefornt control for the off-axis star (B).}
\end{center}
\end{figure}

 As described in previous papers \cite{Thomas14}, \cite{Thomas15}, we identified three main challenges associated with observing double-star (or multi-star) systems --illustrated in Figures \ref{fig:issues1} and \ref{fig:issues2} . (a) Even if we assume the on-axis star is perfectly suppressed or otherwise ignored, and only consider the off-axis star, the off-axis star will often lie beyond the outer working angle (Half-Nyquist frequency) of the deformable mirror (especially for nearby systems such as Alpha Centauri). (b)  Light from different stars is mutually incoherent and must be suppressed independently but simultaneously for each star. (c) Demonstrating the combination of the first two challenges, and doing so in broadband light.

In Thomas et al. 2015 \cite{Thomas15}, we presented a technique called "Super-Nyquist Wavefront Control" (SNWC) which is a solution to the first challenge (limited field of view, or FOV, correctable by a given deformable mirror), and demonstrated computer simulations of SNWC. We also outlined proposed solutions to the other two challenges, which we call, respectively, Multi-Star Wavefront Control (MSWC), and Multi-Star Super-Nyquist Wavefront Control (MSSNWC). 

In the present paper, we go a step further and demonstrate MSWC and MSSNWC with computer simulations. In particular, we present the results of the double star simulation for two different configurations: (a) the separation of the two binaries is in within the correctable FOV of the deformable mirror (b) the separation of the two binaries is larger than the correctable FOV of the deformable mirror. Both configurations come with different challenges. If the binary is tight, the deformable mirror does not have to be exercised in a non-standard way but the flux of the secondary target is high and requires more stroke to cancel. If the separation of the binaries is large, the flux contaminating the dark region is lower but the dark region is in the non-standard controllable zone of the mirror. 
Section \ref{sec:MSWC} describes the Multi-Star Wavefront Control (MSWC) theory and results. Section \ref{sec:MSSNWC} describes the Multi-star Super-Nyquist Wavefront Control (MSSNWC) theory and results. Both configurations are presented in monochromatic light in this paper.

\section{Multi-Star Wavefront Control (MSWC)} 
\label{sec:MSWC}
\subsection{Theory} 
In a well-baffled coronagraphic system, the light from a star on the focal plane can be thought of as consisting of two separate components: (a) diffraction such as Airy rings from the telescope and all other optical components (which are predictable a priori, fixed in time in space, and are fundamentally due to the physics of light); (b) aberrations, or "speckles" from these components (which are random and/or varying in time, and are due to engineering limitations rather than physics). In general, coronagraphs suppress diffraction but not aberrations. An active control system is necessary to suppress aberrations. This same active control is also capable of suppressing diffraction to some degree. Diffraction usually dominates aberrations very close to the star (Airy rings are brighter than speckles), which implies that a coronagraph is useful in that case. However, beyond a certain angle from the star (usually about $10 \lambda/D$), diffraction becomes faint enough that wavefront control systems can usually suppress it without the need for a coronagraph. Furthermore, aberrations often decay with angle slower than diffraction does and thus become dominant at larger angles. In other words, a wavefront control system is both necessary and sufficient to suppress starlight at larger angles from the star. Therefore, our solution is based on wavefront control, rather than coronagraphic suppression of the off-axis star. Note that a simple mask blocking a second star can still be useful as a baffle if the rest of the system is insufficiently baffled.
\begin{figure}[ht]
\begin{center}
   \includegraphics[height=5cm]{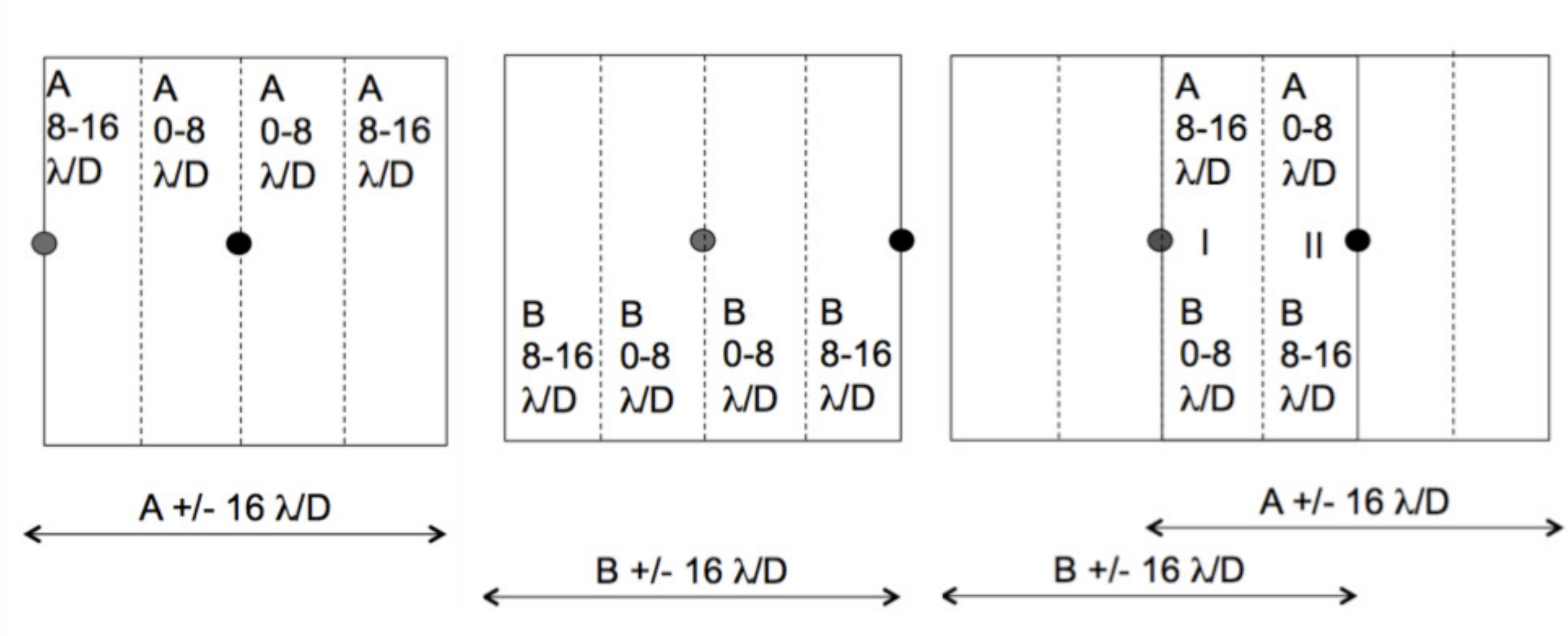}
\caption{\label{fig:MSWC} Principle of Multi-Star Wavefront Control, which uses conventional single-star wavefront control hardware (for this example one 32$\times$32 DM), and a special algorithm. Two stars (black A and grey B) are separated by $16 \lambda/D$. Sub-Nyquist wavefront control regions are shown with respect to star A (left), star B (center), and both stars together (A and B). Vertical dashed lines divide regions showing which DM modes control which star in that region. In regions I or II, independent suppression of both stars can be achieved. In general, one can always find such regions regardless of star separation, the only limitation is the controllable region in a binary system mush be half of that in the single-star system.}
\end{center}
\end{figure}

The key principle of most high-contrast wavefront control methods existing today is to use a Deformable Mirror to diffract a little bit of starlight from the main beam into the science plane region of interest, in such a way as to cause destructive interference with the unwanted speckles initially present there. This of course requires that starlight interfere coherently with the initial speckles in the region of interest. This will always be the case for a single-star target (at least assuming a point-like star and a well-performing system with no technical complications such as poor baffling, unresolved jitter, etc.) However, in the presence of two or more stars, light in any region of interest in general consists of an incoherent combination (both diffraction and aberrations) from all stars. Therefore, in order to remove light from such a region of interest, we must independently and simultaneously suppress each star's speckles by destructively interfering them with only that star's light. In other words, any multi-star wavefront control system must have the property that it removes light for each star independently from the other star(s).

It turns out that this is possible with existing wavefront control hardware and only requires a modification of the software algorithms, albeit a cost of reducing the size of the maximum controllable region of interest. The basic principle is illustrated in Figure \ref{fig:MSWC} and is based on separating DM modes into two independent sets in a way that each set primarily affects the light of one star in the region of interest but not the other. For this example, we consider the case of two stars (black dot A and grey dot B) separated by 16 $\lambda/D$ in the x direction. Figure  \ref{fig:MSWC} (left) shows the full sub-Nyquist control region of a 32 $\times$ 32 deformable mirror out to +/- 16 $\lambda/D$ in both coordinates with respect to star A. Consider two sets of DM modes: the "low-order x" and "high-order x" modes, corresponding to ones whose x-direction spatial frequency content is limited to 0-8 and 8-16 cycles per aperture, respectively. The low order modes primarily affect the regions between 0-8 $\lambda/D$ in the x-direction with respect to each star. Similarly, the high order modes primarily affect the 8-16 $\lambda/D$ region with respect to each star, as shown in Figure  \ref{fig:MSWC}, (left and center, showing the situation with respect to the A and B stars, respectively). Figure  \ref{fig:MSWC} (right) shows a diagram of the superposition of the left and center images. In the region labeled "I", the low-order modes primarily affect speckles of the B star and the high order modes primarily affect the speckles of the A star, so that in that region speckles from both stars are independently controllable with a single DM. A similar situation occurs in region II. Therefore, as long as we limit our region of interest to either I or II, we can achieve independent control of the speckles from both stars. 

It the stars are separated by a distance other than 16 $\lambda/D$, similar regions can be found (as long as the separation is sub-Nyquist -- the super-Nyquist case is treated below). Such regions will have more complicated shapes but will always have maximum total area equal to half of the maximum single-star independently controllable region. This is because the number of degrees of freedom on the DM is fixed, so that doubling the number of stars must necessarily halve the controllable region. We can also generalize this to systems with more than 2 stars -- in that case we can also find regions where light from all stars can be independently controlled with a single DM (or 2 if control of both phase and amplitude errors is desired, just as in the case of single-star wavefront control). In general, the shape of each such region may be complicated and sometimes not even convex or connected, but its total area will in general be 1/N less than the area of a single-star region for that wavefront control system. (This is the price for multi-star wavefront control.)

So far we have only talked about being able to independently control speckles from multiple stars, but mere control does not necessarily imply the ability to find a DM setting where the speckles are actually suppressed. We can demonstrate this solution exists as follows. Most single-star wavefront control algorithms are based on solving a system of linear equations relating DM modes and starlight electric field in the region of interest. As long as we can measure the electric field of each star separately, we can then combine the linear sets of equations for each star into a larger set of linear equations, and simply solve that set. If each star's system of equations is solvable by itself, and each star uses different DM modes, then the combined system must also be solvable. 

A remaining component is the "estimation problem", i.e. the problem of estimating each star's electric field in the region of interest separately (which is one of the necessary parameters in the system of equations for control in the above paragraph). This can be done in a manner analogous to the above paragraph: single-star wavefront control methods usually use the deformable mirror to diffract some trial "probe" light in the region of interest. This modulates the initial speckles there, and this information can be used to estimate their electric fiend by solving another system of linear equations relating the electric field and DM probe shapes. Thus, we have a system of linear equations for each star, which we can combine to yield a solution to the electric field of each star in the region of interest. As long as each of the single-star systems is solvable by itself, and as long as different DM modes affect primarily different stars (i.e. are not degenerate), then the combined system must also be solvable.

\subsection{Results} 
The simulation process is similar to the one described in Thomas et al, 2015 \cite{Thomas15}. In order to create a dark zone around one of the two components of the binary, we consider the correction part of an Electric Field Conjugation algorithm (EFC) \cite{Giveon07} coded in Matlab. We assume that we know the wavefront to be corrected and therefore do not consider the estimation part of EFC as mentioned in the previous section. We added 10nm rms worth of aberrations, following a power law with a coefficient equal to -2. The aberrations were introduced in the pupil plane. In addition, we focus in this paper on monochromatic light. However, the goal is to run the simulations with a 10\% bandwidth, which is computationally expensive. The separation of the star is set to 10 $\lambda$/d (sub-Nyquist) and the region of interest is 2-6 $\lambda$/D. 
\begin{figure}[ht]
\begin{center}
  \includegraphics[height=6cm]{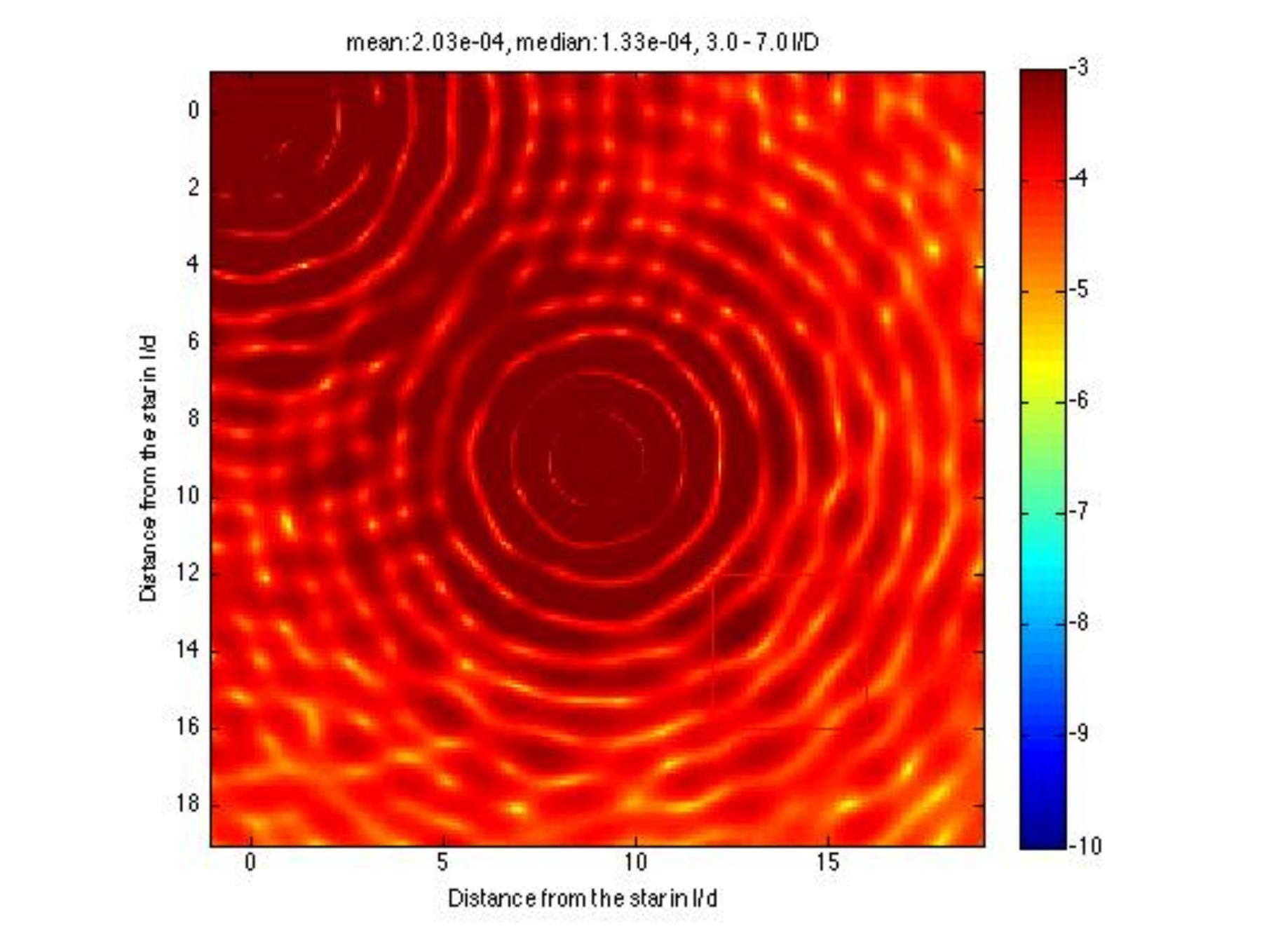}
 \includegraphics[height=6cm]{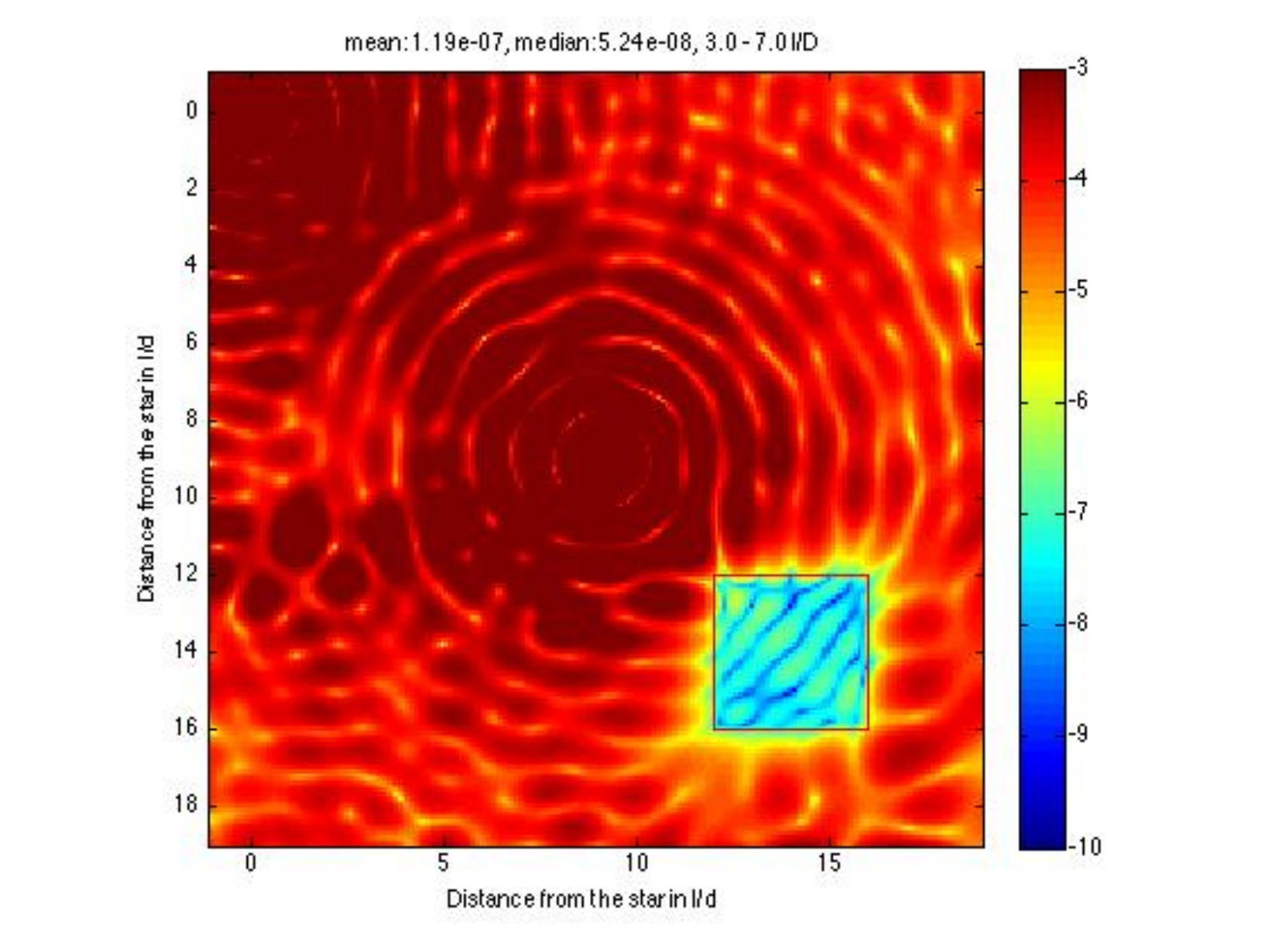}
\caption{\label{fig:MSWC_res} Preliminary proof-of-principle simulation of Multi-Star Wavefront Control, before (left) and after (right) correction. The deformable mirror is a 32 $\times$ 32.  The contrast achieved over a 4 $\times$ 4 area is 5.2$\times 10^{-8}$. A colored version is available electronically.}
\end{center}
\end{figure}

Figure \ref{fig:MSWC_res} shows the results. The contrast was improved from 1.3$\times 10^{-4}$ to 5.2$\times 10^{-8}$.  Note that the performance here is limited by the on-axis diffraction rings and better results are achieved when using a coronagraph to block the on-axis star, which we did for the next section.

\section{Combining SNWC and MSWC: MSSNWC } 
\label{sec:MSSNWC}
\subsection{Theory}
In the previous section, we described how MSWC can remove light for the case that the separation between two stars is sub-Nyquist. In Thomas et al 2014 \cite{Thomas14} and Thomas et al 2015 \cite{Thomas15}, we demonstrated SNWC, a method to suppress starlight from a star beyond the DM Nyquist frequency. In this section we demonstrate how these two techniques can be combined theoretically and then demonstrate an example by a computer simulation.

\begin{figure}[ht]
\begin{center}
   \includegraphics[height=12cm, angle=270]{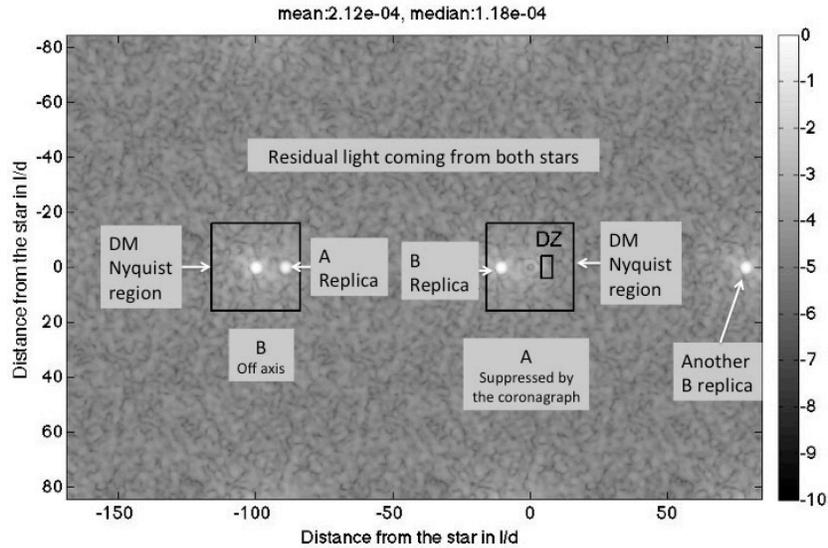}
\caption{\label{fig:MSSNWC_solution} Schematic of the MSSNWC scenario: 2 stars separated by 100 $\lambda$/d. The on-axis star A is now blocked by a simple apodized Lyot coronagraph. The star B is replicated due to the grid off the DM. The grid is due to the 32 $\times$ 32 actuators from the DM.}
\end{center}
\end{figure}

Figure \ref{fig:MSSNWC_solution} shows a field of view containing two stars: a coronagraphically suppressed star A (centered inside the black square on the right), and star B (centered in black square on the left). The back squares represent the DM sub-Nyquist region, and the small "DZ" rectangle represents the region of interest, or "dark zone" in the planetary system around star A. The dark zone consists of contribution of light from both stars. If star B was in the sub-Nyquist region of star A, we could just apply MSWC to remove both stars' light from the DZ, but star B is too far away for MSWC. In Thomas et al. 2014 and 2015 \cite{Thomas14}, \cite{Thomas15}, we described how SNWC is essentially equivalent to conventional (sub-Nyquist) wavefront control, but applied with respect to one of the diffraction orders of the star caused by a mild grating in the pupil plane. The segmentation of segmented telescopes or segmented DMs already acts as an appropriate grating, or the "quilting" pattern left on most DMs as part of their fabrication process can suffice. If neither of those are present, a mild grating \cite{Guyon12}, \cite{Bendek13} can be deliberately installed in the system. In any case, if this grating generates a (faint) diffracted copy of star B inside the sub-Nyquist region of star A, then we can now use this diffracted copy of star B as its "proxy" and apply MSWC on this proxy and star A. This essentially reduces the super-Nyquist multi-star problem to the already solved MSWC, in exactly the same way as SNWC reduces the single-star super-Nyquist problem to the single-star sub-Nyquist problem.

Note that Figure \ref{fig:MSSNWC_solution} shows a monochromatic rather than broadband simulation. In broadband light, the diffracted copy of star B will be chromatically elongated along a line pointing to star B, which would seem to pose a complication and unique challenge in broadband. However, all other light from star B (speckles, other DM modes) will also be chromatically elongated in a similar fashion. Thus, the elongation of the proxy star should help, rather than hinder, broadband control in that region. Indeed, as we already showed in the computer simulations in \cite{Thomas15}, SNWC works fairly fell in broadband, at least under certain assumptions. Although we have not yet demonstrated broadband operation in general, the above arguments suggest that broadband challenges are entirely a function of stellocentric angle and not whether super-Nyquist or sub-Nyquist control is used. For example, SNWC at 24 $\lambda/D$ with a 32 $\times$ 32 DM (where the sub-Nyquist boundary is 16 $\lambda/D$) should not cause any unique broadband challenges beyond those already associated with performing conventional (sub-Nyquist) wavefront control at 24 $\lambda$/D with a 64 $\times$ 64 DM. In the same way, we expect that SNMSWC will not cause any unique broadband challenges.

\subsection{Results}
For this section, we simulated a portion of an Airy pattern resulting from a normal circular aperture at 25 $\lambda / d$. For this preliminary simulation, we use an Apodized Lyot Coronagraph to block the light coming from the center star and we used a grid of dots with a frequency equal to 32 $\lambda / d$, which corresponds to the regular pattern that MEMS (Micro-Electro-Mechanical system) create. Because they have a fixed number of actuators the frequency that the MEMS can create has to be a multiple of the number of actuators. 
The simulated dots are the size of a pixel and therefore depend on the resolution of the simulation.  The amount of aberrations included in the pupil plane was 10nm with a power law of -2.
The preliminary results presented here are in monochromatic only. The ultimate goal is to repeat the experiment in 10\% band but the computation was very expensive in order to get a goal sample of the bandwidth. We moved the code over to the super computer facilities at NASA Ames and will present results in a future paper.

\begin{figure}[ht]
\begin{center}
  \includegraphics[height=7cm]{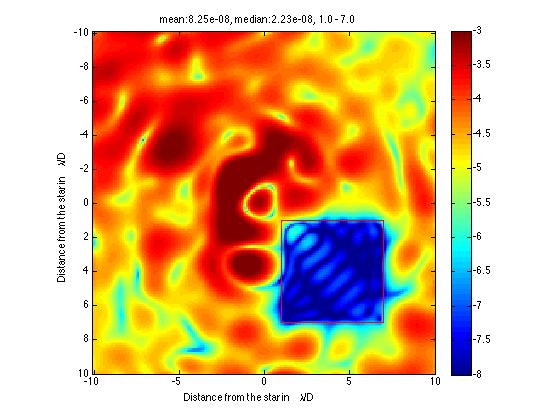}
\caption{\label{fig:MSSNWC} Results of the MSSNWC method, combining the SNWC and the MSWC. The separation of the two stars is 25 $lambda$/d and is located in the upper left corner in the above image. The on-axis star is blocked by an apodized Lyot coronagraph and the grid has a dots frequency of 32. A colored version is available electronically.} 
\end{center}
\end{figure}

The contrast achieved here is $2 \times10^{-8}$ with a 6 $\lambda/d \times 6 \lambda/d$ region. If we adjust the dark zone region to with a 4 $\lambda/d  \times 4  \lambda/d$ region, the contrast reached is less than $10^{-8}$. Note that these results are preliminary.

\section{Conclusion and Applications}
We presented the basic principles and first computer simulations of methods that enable high-contrast imaging of exoplanets and disks in multi-star systems. This is particularly important because most Sun-like and earlier stars are in multi-star systems. Of special note is Alpha Centauri, which is not only the closest star system to Earth, but also ~2.5x closer than the next closest non-M-dwarf star. It would be the most favorable target (by a large margin) for any direct imaging telescope capable of suppressing light from a binary. 

Our methods are called Multi-Star Wavefront Control (MSWC) and Multi-Star Super Nyquist Wavefront Control (MSSNWC), depending on whether the separation between the two stars is sub-Nyquist or super-Nyquist. Both of them rely on a single-star coronagraph and suppress the other star(s) with wavefront control only because the dominant source of noise from the other stars is aberrations and not diffraction. Both methods use existing, high technology readiness single-star imaging systems with no required changes, and only use special software algorithms. Suppression of multiple stars is achieved simply by using different modes of the deformable mirror on different stars. This technique requires reducing the maximum area of the high-contrast region of interest, which can be enlarged by using a DM with more actuators.
 
\acknowledgments     
This work was supported in part by the National Aeronautics and Space Administration's APRA (Astrophysics Research and Analysis) program through solicitation NNH13ZDA001N-APRA at NASA's Science Mission Directorate, and NASA Ames Research Center. It was carried out at the NASA Ames Research Center. Any opinions, findings, and conclusions or recommendations expressed in this article are those of the authors and do not necessarily reflect the views of the National Aeronautics and Space Administration.

\bibliography{MSSNWC_spie}   
\bibliographystyle{spiebib}   

\end{document}